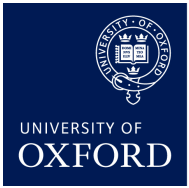 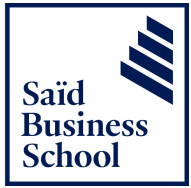

# Uniqueness Bias:

# Why It Matters, How to Curb It

August 2024

Bent Flyvbjerg, Alexander Budzier, M. D. Christodoulou, and M. Zottoli

University of Oxford

IT University of Copenhagen

## Working paper | 2023-24

# Abstract

The paper explores "uniqueness bias," a behavioral bias defined as the tendency of planners and managers to see their decisions as singular. For the first time, uniqueness bias is correlated with forecasting accuracy and performance in real-world project investment decisions. We problematize the conventional framing of projects as unique and hypothesize that it leads to poor project performance. We test the thesis for a sample of 219 projects and find that perceived uniqueness is indeed highly statistically significantly associated with underperformance. Finally, we identify how decision makers can mitigate uniqueness bias in their projects through what Daniel Kahneman aptly called "decision hygiene," specifically reference class forecasting, premortems, similarity-based forecasting, and noise audits.

# Introduction

Behavioral economics is a successful research paradigm across the social sciences. It has several Nobel Prizes to its name and cultivates the kind of theoretical debate that helps consolidate a field in academia (Kahneman and Klein 2009, Gigerenzer 2018, Flyvbjerg 2021). Studying biases in "real-world" decision making (as opposed to lab settings) is a growing, yet underdeveloped, field (Toplak et al. 2017). In this paper, we return to one of the first real-world settings studied by Kahneman and Tversky (1977), namely project investment decisions. Kahneman and Tversky's early work was foundational for how to think about projects. Nevertheless, a systematic study of behavioral theory is largely absent from the project management literature. A recent, comprehensive review concluded that "few [studies] were found which recognize behavior as an underlying construct that influences the outcome of technical project processes" (Ramirez 2021: 24). Some project management scholars even advise against understanding their field in terms of behavioral economics (Flyvbjerg et al. 2018, Flyvbjerg 2018).

This state of affairs differs from that of other parts of the management and social science literature. The situation is easy to understand, given project management's historical roots in engineering, where projects conventionally have been seen to be more about things than about people. In this paper, we continue the emergent work of exploring the relevance of behavioral science to project management scholarship and practice. We define and explore a behavioral bias called "uniqueness bias," which we argue is particularly pertinent to project management. We explain the bias in terms of (a) individual and collective decision making (Buehler et al. 2005, Fiedler et al. 2019, Blanchard et al. 2020), (b) extreme cases (Plonsky and Teodorescu 2020, Paine et al. 2020, Kirshner 2021, Dannals and Oppenheimer 2022), and (c) the role of learning (Leisti and Häkkinen 2018).

# What Is Uniqueness Bias?

Psychologists originally identified uniqueness bias as the tendency of individuals to see themselves as more singular than they actually are, e.g., singularly healthy, clever, or attractive (Suls and Wan 1987, Suls et al. 1988, Goethals et al. 1991, Rose et al. 2017). Uniqueness bias is related to, but distinct from, other well-studied biases, such as overconfidence bias and the illusion of learning. Overconfidence bias – whether as overestimate, overplacement, or overprecision – directly relates a person's judgment to that of others (Moore and Healy 2008). The illusion of learning is the misplaced confidence in having learned from others and overcoming their mistakes when, in fact, you have not (Metcalfe 1998). Uniqueness bias is different in that it involves a judgment about the degree to which something is perceived to be singular, of little or no relevance to anything else, and vice versa. In project management, the term "uniqueness bias" was first used in print by Flyvbjerg (2014: 9), who defined it as the tendency of planners and managers to see their projects as singular, i.e., as one of a kind. The term was developed further in Flyvbjerg (2021), while the present paper is the first full treatment of the subject in the context of decision making and project management, including the first empirical test against real-world project investment decisions.

In projects, uniqueness bias may be a non-deliberate cognitive bias, as in its original formulation by psychologists. Or it may be a strategic bias used deliberately by planners and managers to lure people to their projects, for instance, based on the reasoning that something presented as unique and new is more likely to attract support and funding than something that has been seen and done before (Flyvbjerg 2021). Most projects do not advertise themselves as copies of something, even if they are. E.g., Bumble, a dating application, did not market itself as a knockoff of Tinder made by former Tinder employees, but quite the opposite. Uniqueness bias can be seen as the result of judgments based on representativeness (Kahneman and Tversky 1972) and similarity (Read and Grushka-Cockayne 2010).

Uniqueness bias is a general bias, but it is particularly interesting as an object of study in project planning and management because project planners and managers are systematically primed by their professional organizations to see projects as unique, including at the basal level of defining what a project is. For instance, the standard definition of a project according to the biggest professional organization in the field, the US-based Project Management Institute (PMI), directly emphasizes uniqueness as one of two defining features of what a project is: "A temporary endeavor undertaken to create a *unique* product, service, or result" (PMI 2021: 4, italics added). For a long time, the UK-based Association for Project Management similarly defined a project as a "*unique*, transient endeavor" (APM 2012, italics added) but recently thought better of it and dropped "unique" from their definition (APM 2019).

Academics, too, define projects in terms of uniqueness. Indeed, the very first study of projects as a management problem (Gaddis, 1959: 90) identified the finite duration of projects as a "*unique* aspect of the project manager's job" (italics added). Hirschman (1967: 186) similarly concluded that each project he studied for his classic book, *Development Projects Observed*, turned out to represent "a *unique constellation* of experiences and consequences, of direct and indirect effects" (italics in the original; see also Flyvbjerg 2016).

More recently, Turner and Müller (2003: 7) defined a project as follows: "A project is a temporary organization to which resources are assigned to undertake a *unique*, novel and transient endeavour managing the inherent uncertainty and need for integration to deliver beneficial objectives of change" (italics added). Other examples of writers who define projects in terms of uniqueness are Baccarini (2004), Chapman (1998), Conroy and Soltan (1998), Czuchry and Yasin (2003), Fox and Miller (2006: 3, 109), Gareis (1991), Grün (2004: 3, 245), Jaafari (2007), Merrow (2011: 161), and Perminova et al. (2008).

We observe that the understanding of projects as unique appears to be deeply entrenched in the project management profession, the literature, and individual project managers and scholars. We maintain that this is unfortunate for two reasons. First, taken to their logical consequence, the definitions of projects as unique are nonsensical in the sense that if projects were indeed truly unique then the only thing they would have in common is that they have nothing in common.[1] This does not appear to us a productive point of departure for developing a deeper understanding of projects, but quite the opposite. Second, and more importantly, the conventional definitions contribute to uniqueness bias, in the grip of which project planners and managers tend to see their projects as more singular than they actually are. This is reinforced by the fact that new projects often use non-standard technologies and designs. As with other behavioral biases, uniqueness bias results in a distorted view of reality that underestimates risk and overestimates opportunity, thus negatively impacting the quality of decisions.

Specifically, uniqueness bias will likely impede project managers' learning because they will think they have nothing (or little) to learn from other projects if their own is unique, even though the benefits of learning on decisions are well established (Leisti and Häkkinen 2018). We test this thesis below. Grün (2004: 245) sees the world from the "no-learning" perspective when he approvingly quotes Magee (1982: 4) for saying that big, complex projects "are so different from one another that no learning factor or experience curve can yet be applied to them." Below, we argue this is wrong.

---

[1] The authors wish to thank Jens Schmidt at The IT University of Copenhagen for help with arriving at this formulation.

We further contend that only projects with a significant learning factor and experience curve are worth doing and that many such projects exist. If projects with no learning factor or experience curve existed, they would be projects to avoid doing because they would likely go very wrong for lack of anything to build on in terms of learning and experience.

Similar to Grün, Jennings (2012: 77-78) quotes the British Olympic Association (2000: 48) to argue the common view that "inherent geo-social differences of all major cities" are so big that each Game must be considered unique and that learning in terms of cost, therefore, "cannot be accurately extrapolated from the experiences of other cities." But this is equivalent to arguing, e.g., that subways in cities must be considered unique because each city is different, which would be a recipe for disaster in terms of decision making and project management. Jennings's view is common, however, resulting in ruinously high cost overruns for hosting the Olympic Games (Flyvbjerg et al. 2021, Budzier and Flyvbjerg 2024). Decision makers and scholars use the uniqueness argument to explicitly argue against the possibility of learning from others and, therefore, implicitly against effectively improving decisions in the long run.

Investing in a project is typically not the decision of a single individual but of a group of individuals belonging to one or more organizations. We might, therefore, perhaps expect a wisdom-of-crowds effect that would mitigate individual biases, including uniqueness bias, in such a decision. Unfortunately, the extant literature does not support this expectation. Groupthink (Janis 1983) is a well-documented issue in collective decision making, as is the tendency of collectives to make riskier decisions than individuals (Blanchard et al. 2020). Buehler et al. (2005) found that task completion estimates made by groups are more optimistic than estimates made by individuals, an effect they ascribe to the tendency of groups to plan for success. Teams also show myopia and overconfidence in their abilities compared to others (Radzevick and Moore 2008). It appears that despite a broader set of individual experiences, project decisions often fall prey to bias, even under collective decision making and even when the collective is made up of experts (Bazerman 2001). This applies to bias in general and, therefore, most likely, also to uniqueness bias.

A project may, of course, be unique in its specific geography and time. For instance, California has never built a high-speed rail line before, so in this sense, recent efforts to build high-speed rail between Los Angeles and San Francisco, or some part of this, may be considered a unique project (Flyvbjerg and Gardner 2023). However, the project is only unique to California and, therefore, not truly unique. Dozens of similar rail projects have been built around the world, with data and lessons learned that would be highly valuable to California in terms of costs, schedule, contracting, procurement, revenues, impacts, etc., should the Californian high-speed rail authorities decide to systematically benefit from this.

In that sense, projects are no different from people. A quote, often ascribed to cultural anthropologist Margaret Mead, captures the point well: "Always remember that you are absolutely unique. Just like everyone else." Each person is self-evidently unique; just ask their parents. However, each person also has lots in common with other people; just ask doctors. The uniqueness of people has not stopped the health sciences and health professions from making progress based on what humans have in common. The problem with project management is that uniqueness bias hinders such learning across projects because project managers and scholars are prone to a "localism bias," which we define as the tendency to see the local as global due to availability bias for the local. Following this argument, the root cause of uniqueness bias is localism bias, and the latter bias explains why local uniqueness is so easily and often confused with global uniqueness. In many projects, planners and managers do not so much as look outside their project because "our project is unique," which is a mantra you hear over and over in projects and which it is surprisingly easy to get project managers to admit to. Or perhaps it is not so surprising once you understand the workings of uniqueness bias and other behavioral biases.

Even Daniel Kahneman, by many considered the leading intellectual light of behavioral science, was tripped up by uniqueness bias in an episode he calls "one of the most instructive experiences of my professional life" (Kahneman 2011: 245-247). Kahneman and several colleagues planned to write a textbook together. They agreed it would take roughly two years, with the lowest estimate at one and a half and the highest at two and a half years. Only one member of the group had considerable experience with writing textbooks. Kahneman asked this person how long such books usually take, and the expert said he could not recall any project taking less than seven years. Worse, about 40 percent of projects are never finished. Kahneman and his colleagues were briefly alarmed. This was much longer than they had estimated, and they had not considered the possibility they might fail. "We should have quit," Kahneman concluded years later. Instead, after a few minutes of debate, they carried on as if nothing had happened because the forecast "seemed unreal" (p. 246). Their project felt different, according to Kahneman. It always does. "This time is different" is the motto of uniqueness bias. The textbook was ultimately finished *eight* years later. If the greatest student of cognitive bias can be suckered by uniqueness bias, it is no wonder that project managers are vulnerable, too.

In sum, uniqueness bias feeds well-established behavioral phenomena like the inside view, base-rate neglect, and optimism bias, which all feed the underestimation of risk. Underestimation of risk distinguishes uniqueness bias from statistical noise. If uniqueness bias resulted in noise, then overestimates and underestimates of risk would be roughly equally frequent and would cancel out each other, more or less (Kahneman et al. 2021). But that is not what happens under uniqueness bias. Here, risk is underestimated, as further documented below, leading to a double whammy of underestimated cost risk compounded by underestimated benefit risk. Therefore, uniqueness bias is a true bias, making project teams take risks they would likely not have accepted had they known the real odds. Good project leaders do not let themselves be fooled like this. They know that those who say projects are unique are mostly wrong. Projects are often unique locally, yes. But to be locally unique is an oxymoron. This, however, is the typical meaning of the term "unique" when used in project management. It is a misnomer that undermines project performance and, therefore, project management itself. Truly unique projects are rare. We have much to learn from other projects, always. And if we do not learn, we will not succeed with our projects.

## Does Uniqueness Bias Impact Project Performance?

Uniqueness bias caused Kahneman and his co-authors a 300 percent schedule overrun on their book project, as described above. We decided to try and get beyond anecdotal evidence and test the relationship between uniqueness bias and project performance more systematically.

For this, we established a sample of 219 IT projects from two sources. First, we used our professional network to connect us with organizations willing to share their data on project performance. Second, we cold-called other organizations, which we thought might have suitable data, and directly solicited their participation. In total, we contacted 365 organizations, of which 34 had relevant data they were willing to share. We included only organizations with a formal, written record of estimated and actual project performance. Further, we accepted only projects that were completed to allow ex-post analysis. Each organization was asked to provide data for as many projects as possible that met these criteria and had been completed within the last five years at the time of data collection. We included all types and sizes of IT projects to avoid possible biases associated with restricting the sample to particular project types or arbitrary cut-off values regarding size. In total, this resulted in quantitative performance data for 1365 projects. We then asked each of the 34 organizations to allow us to interview team leaders on the 10 to 15 most recent of their projects for which we had already collected performance data to collect further subjective data on perceived uniqueness. We asked for the 10 to 15 most recent projects to randomize the types of projects and interviews without sampling on project performance. It was possible to collect further data on perceived uniqueness through interviewing for 219 projects. This is a response rate of 43 to 64 percent. The main reasons for non-responses were that leaders had left the organization or they had worked for a third party during the project. The 219 projects we interviewed were located in Europe, North America, the Middle East, Africa, Asia, and Australasia. They ranged in size from a total investment of US dollars 77,000 to 4.5 billion (2022 dollars). Six projects were public sector, 213 private. The projects were completed from 2005 to 2014.

Data collection probably introduced a selection bias in the sample, to the extent that organizations that keep systematic data records and are willing to share these are likely to be more mature and better managed, with better project performance, than average. Therefore, results from the sample are likely to show better performance than results from the project population. This should be kept in mind when interpreting the findings below.

For each of the 219 projects, we asked managers how unique they deemed their project to be by asking how much they agreed/disagreed with the following statement: "This project is unique, and therefore it is difficult to compare with other projects." Responses were measured on a Likert scale from one to ten using descriptive anchors, with one signifying "The project was not unique at all," four described as "Project had a few unique aspects to it; but not fundamentally different from other projects, "seven as "Project was unique to parts of the team, users, e.g., experienced vendor, or project leaders; difficult to compare to other projects," and ten as "truly one of a kind project, seen as unique by team, vendors, and clients." The values of five to six are in the middle range of the scale. Figure 1 shows the frequency distribution of answers.

*Figure 1: Frequency distribution of answers to the question of how unique IT project managers deemed their projects to be on a Likert scale from one to ten, with one signifying "not unique at all," ten signifying "truly one of a kind, seen as unique by team, vendors, and clients," and values in between denoting various degrees of uniqueness. The question was asked for a total of 219 IT projects.*

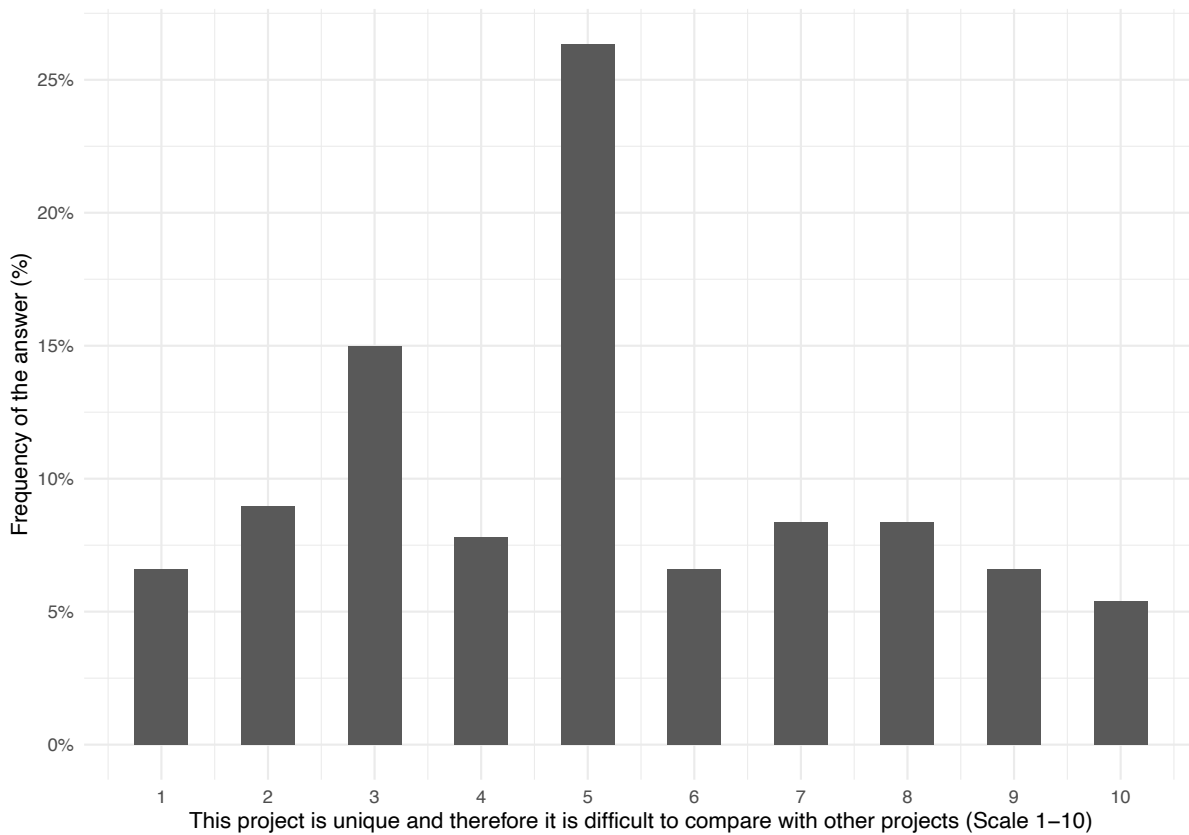

After having measured the degree of perceived uniqueness for each project, we used linear and logistic regression to test the association between perceived uniqueness and performance, with the latter measured on four separate variables: (a) cost overrun (N = 167), (b) schedule overrun (N = 108), (c) benefit overrun (N = 38), and (d) percentage of extreme values of cost overrun (N = 167). We defined overrun in the conventional manner as the ratio of actuals divided by estimates expressed in percent. Estimates were taken from the final investment decision. Actuals were measured after project completion.

Figure 2 shows the correlation between perceived uniqueness and percentage cost overrun in real terms. We see that an increase in perceived uniqueness correlates positively with an increase in cost overrun. Specifically, a one-point increase on the ten-point Likert scale associates with, on average, a five percentage point increase in cost overrun, and the relationship is statistically significant.[2] The results further show that truly one-of-a-kind projects (ten on the Likert scale) should, on average, expect a 45 percentage point higher cost overrun in real terms than not-unique-at-all projects (one on the Likert scale). This is a large and significant difference. The data clearly support the thesis that project leaders' viewing their projects as unique correlates with underperformance. Segelod (2018: 166-168) argues that the more unique a project seems to be to its cost estimators, the higher the cost overrun will be, and gives anecdotal examples of this. The above analysis is the first systematic study of the problem that we know of, and it proves Segelod's anecdotal hunch to be correct.

---

[2] Robust linear regression with sandwich estimator, p = 0.01, estimated intercept = 0.89, coefficient mean estimate = 0.049, sandwich corrected 95-percent CI of the coefficient effect estimate ranges from 0.011 to 0.087, sandwich corrected z-value = 2.500, sandwich corrected p-value = 0.012.

For schedule overrun, we found that an increase in perceived uniqueness does not correlate with schedule overrun (Kendall's tau = 0.05, p = 0.524). For benefit overrun, too, results were non-significant (Kendall's tau = 0.11, p = 0.423), likely because fewer data were available. For schedule and benefit overrun, the collection of more data is an area for further research.

*Figure 2: The relationship between perceived uniqueness and cost overrun is positive and statistically highly significant. The solid line is the linear fit (p = 0.003). A one-point increase in perceived uniqueness correlates with a five percentage point increase in cost overrun. Truly one-of-a-kind projects (ten on the Likert scale) should expect a 45 percentage point higher cost overrun in real terms than not-unique-at-all projects (one on the Likert scale). The data clearly support the thesis that viewing a project as unique is likely to lead to underperformance.*

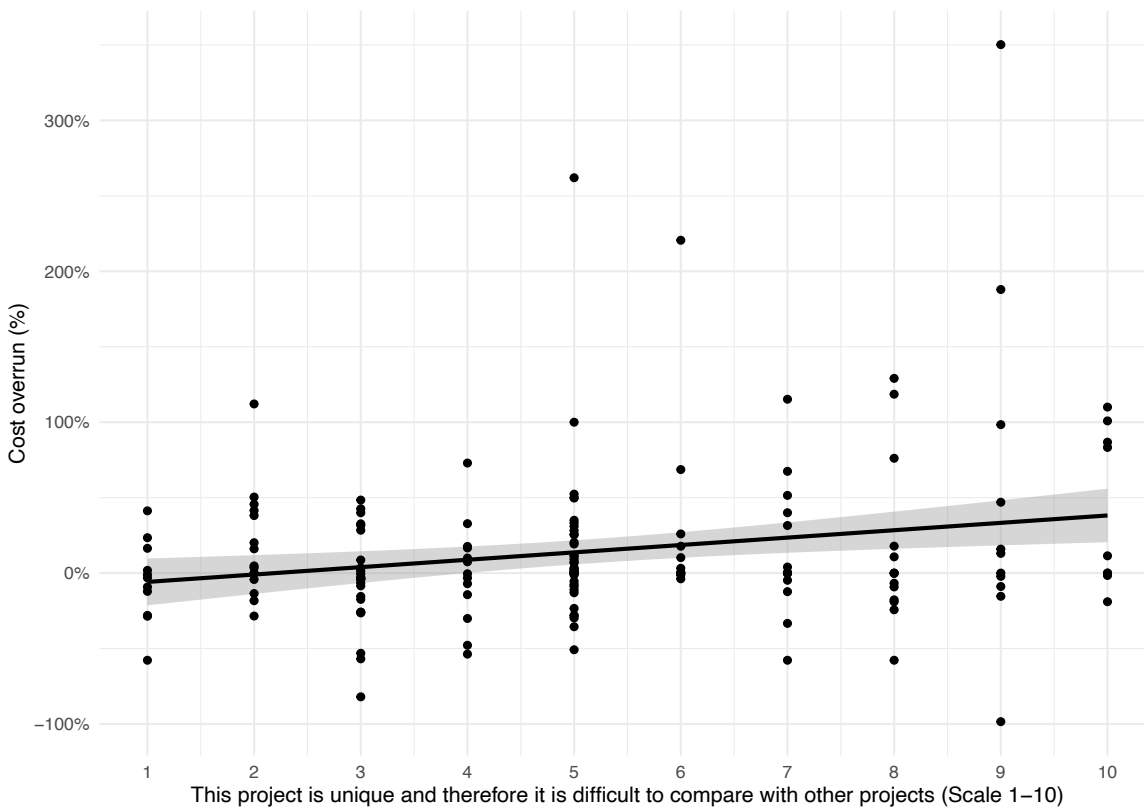

We further investigated the association between project size and perceived uniqueness. It could be argued that truly unique projects might be likely to require more effort in terms of money and time and thus would tend to be larger in those terms. We measured size in four ways: estimated and actual cost and estimated and actual duration, respectively. We found no statistically significant relationship regarding size using any of these measures.[3] Further tests showed no statistically significant association between results and geography (p = 0.99780), public vs. private ownership (p = 0.99), and year of completion (p = 0.46).

---

[3] The statistical tests used were (a) estimated cost: Kendall's tau = 0.045, p-value = 0.4474; (b) actual cost: Kendall's tau = 0.019, p-value = 0.711; (c) estimated duration: Kendall's tau = -0.040, p-value = 0.47; (d) actual duration: Kendall's tau = -0.023, p-value = 0.6651.

Building on studies by Mandelbrot (1997) and Taleb (2007, 2020), we previously found that the prevalence of extreme values (sometimes also called fat tails or outliers) may be more significant than averages in terms of project performance. So far, we have documented this hypothesis for information technology and the Olympic Games, but preliminary research indicates it applies to other project types as well (Flyvbjerg et al. 2022, Flyvbjerg et al. 2021, Flyvbjerg and Gardner 2023).

The study of extreme values in behavioral economics has experienced a boost of late. Research has found that extreme and rare events are overweighted in descriptions and underweighted in experience (Ludvig et al. 2014, Yechiam et al. 2015) and are often excluded from decision making (Dannals and Oppenheimer 2022). Other studies investigated how risk-taking behavior changes after near-miss events (Paine et al. 2020, Kirshner 2021) and forgone knowledge of disasters (Plonsky and Teodorescu 2020). Feduzi and Runde (2014) explored the relationship between extreme values and unknown unknowns, plus the effects on decision biases arising from considering too narrow a range of potential future states (excluding extremes), driven by confirmation bias. In the present study, we investigated the relationship between uniqueness and extreme decision outcomes.

We tested the association between perceived uniqueness and project performance, with the latter now measured by the prevalence of extreme values. Extreme values were measured in the conventional manner, using Tukey's definition of outliers as observations that are 1.5 interquartile ranges or more away from the third quartile, which for cost overrun translates into values of +76 percent or more, which account for 16 percent of all observations in the sample. This indicates a much more fat-tailed distribution of cost overrun than the Gaussian, which would have only 0.7 percent extreme values measured in the same manner. In other words, the actual tail is 23 times fatter than the Gaussian tail.

Figure 3 shows the association between perceived uniqueness and frequency of extreme cost overruns. Again, we see a positive association, which is statistically significant (p = 0.001).[4] A truly-one-of-a-kind project (ten on the Likert scale) has an average risk of 37 percent of extreme cost overrun (≥ 76 percent), compared with an average risk of only one percent for a not-unique-at-all project (one on the Likert scale). Again, this is a large and significant difference. Interestingly, we see that the probability of big cost blowouts accelerates in a non-linear fashion precisely at the level of the Likert scale (five to six), where projects start to be deemed dominantly unique. We conclude once more that the data support the thesis that viewing your projects as unique is associated with underperformance. Similar tests were carried out for schedule and benefits, but they proved statistically non-significant, likely because of small sample sizes. Here, too, collecting more data is an area for further research.

---

[4] Logistic regression, intercept -5.24 with z-value = -5.11, p-value < 0.001; and coefficient mean estimate = 0.44 with z-value = 3.27, p-value = 0.001.

*Figure 3: The association between perceived uniqueness and the probability of an extreme cost overrun (defined as ≥ 76 percent, see main text) is positive and highly statistically significant (p < 0.001, logistic regression). The shaded area is the 95-percent confidence interval. Truly-one-of-a-kind projects (ten on the Likert scale) have a 37 percent risk of extreme cost overrun, compared with one percent for not-unique-at-all projects (one on the Likert scale), with no overlap of confidence intervals.*

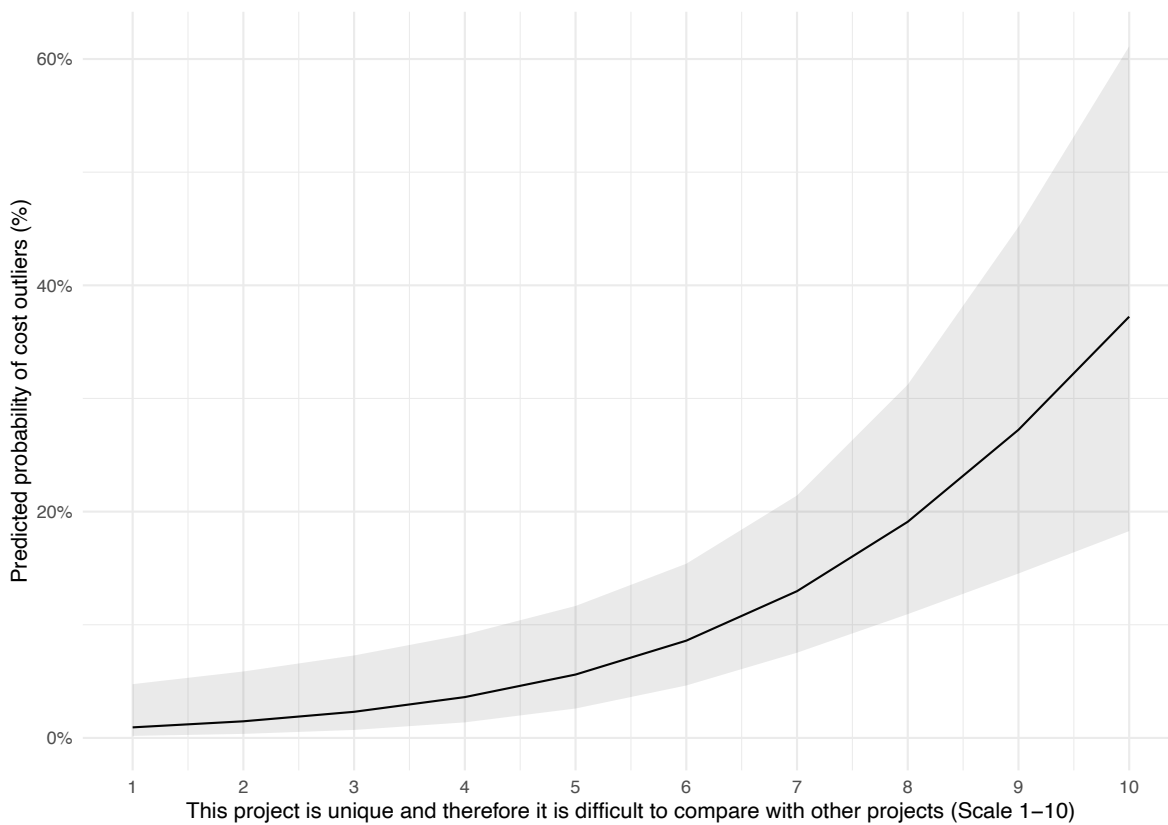

Furthermore, we identified project characteristics that correlate with perceived uniqueness, shown in Table 1.[5] Projects that were complex, political, had many unknowns, shifting requirements, and used an agile approach correlated positively and highly significantly with uniqueness (p ≤ 0.001 and p = 0.005, respectively). Neither of these characteristics had a statistically significant effect above and beyond perceived uniqueness in our logistic regression model to explain extreme overruns. However, the characteristics could reliably identify projects likely to suffer from uniqueness bias and thus underperform.

---

[5] We tested 35 additional project characteristics found in the project management literature. The complete list of characteristics is included in the Methodological Appendix.

*Table 1: Project characteristics that correlate with uniqueness at a high level of statistical significance (Kendall's tau, p ≤ 0.01). Projects that were complex, political, had many unknowns, and shifting requirements correlated with uniqueness at a high level of statistical significance.*

| Project characteristic | Correlation with uniqueness | p-value |
|---|---|---|
| The solution requires complex internal processing in the system | 0.24 | <0.001 |
| The initial plan and business case showed political intent instead of the most likely outcome | 0.19 | 0.001 |
| There were a large number of unknowns in the solution design | 0.19 | <0.001 |
| The project had competing and shifting requirements | 0.17 | <0.001 |
| The project followed an incremental approach | 0.14 | 0.005 |

Finally, we asked managers to what degree they agreed with the question, "If we had known the real costs of the project, it would not have been built," and we tested the correlation of their answers with the degree of perceived uniqueness for their projects. Again, we found a statistically significant positive correlation (Kendall's tau = 0.15, p = 0.018), meaning that with increasing uniqueness – and correspondingly increasing cost overruns and risk of extreme cost blowouts, as we saw above – more managers tended to agree with the question and express a form of decision regret. According to the data, the type of project manager who says, "Our project is unique," therefore, tends also to be the type of manager who says, "If we knew the real costs, nothing would ever get built," and, more importantly, they tend to be the type of manager found on projects with big cost overruns. Therefore, if you are the owner or funder of projects that employ such managers, you might want to retrain or replace them to protect your project and your funds. We explain how below.

## Perceived v. Actual Uniqueness

It should be noted that the above conclusions are based on *perceived* uniqueness (we asked managers how unique they *think* their projects are). The question, therefore, remains what the relationship is between perceived and actual uniqueness and whether it matters. To answer this question, for the 59 projects with a perceived-uniqueness score of seven or higher, we compared the functional scope description and start dates with other projects in our database to judge whether the 59 projects were indeed unique. We found precedence of similar projects in the same organization or industry for all these projects, meaning that none of the projects could be considered truly unique. To illustrate, five of the 59 projects were regulatory compliance projects in banks. We established that each of the banks had completed similar regulatory projects before *and* that every other bank in their jurisdiction was working to introduce the same type of regulation at the same time, based on which we concluded the five projects were not unique, not even close.

Furthermore, we identified all projects with a score of nine or ten on the Likert scale, 23 projects in total, of which 18 were complex legacy system replacements, two were data security projects, two were IT infrastructure upgrade projects, and one was the digitalization of the customer support function for a big multinational. For each, we established whether projects were truly first-of-a-kind. We did this by first developing a short description of the project's scope. We then again compared this with the scope of projects in our database. If we could find one or more projects with similar scope that were implemented at the same time or earlier, we ruled out that the project was truly unique. We found similar, earlier projects for each of the 23 projects with the highest uniqueness ratings. Each was, therefore, at the most locally unique. Localism bias, as described above, i.e., thinking the local is global, had tripped up the managers, leading to uniqueness bias.

On that basis, we conclude from the available data that (a) many more projects are perceived as unique than are actually unique; (b) perceived and actual uniqueness are uncorrelated, i.e., projects perceived as unique are typically not actually unique; (c) perceived uniqueness is what matters to project performance, because project managers' behavior is based on this, e.g., managers will think there is nothing to learn from other projects if they believe their project is unique, whether this is true or not, and lack of learning will likely hamper project performance; and (d) a project that is actually unique would likely also be perceived as unique.

In sum, the data show that the perceived uniqueness of projects matters, and the stronger it is, the more likely the projects are to underperform. Managers who regard their project as unique are, therefore, a liability. This makes it extra unfortunate that professional organizations and the project management literature so actively and explicitly encourage managers to see their projects as unique.

## Do Truly Unique Projects Exist?

If a project was ever truly unique, it must be the Apollo program, which landed the first human on the moon. This had never been done before, so how could it not be unique? Or so people argue. Fox and Miller (2006: 3, 109), for instance, describe the Apollo, Polaris, Atlas, Minuteman, and similar large-scale ventures as "unique" and "first of their kind." Interestingly, however, the NASA leadership at the time of Apollo sharply disagreed with such thinking. They deplored those who saw the Apollo program "as so special – as so exceptional" because such people did not understand the reality of the program and, therefore, placed it at risk. NASA insisted, by contrast, that "the basic knowledge and technology and the human and material resources necessary for the job [i.e., landing a person on the moon] already existed" from hundreds of previous US and Soviet spaceflights, so there was no reason to reinvent the wheel (Webb 1969: 11, 61). That is how intelligent leaders deal with uniqueness bias. They see it for what it is, a behavioral fallacy, and avoid it.

Or take nuclear decommissioning. One day, one of us (Bent) received a call from a Swedish top civil servant in charge of decommissioning nuclear power. He needed a reliable estimate of how much it would cost to (a) decommission Sweden's fleet of nuclear power plants, which would take decades, and (b) safely store nuclear waste, which would last centuries. Sweden's nuclear industry would be asked to pay into a fund to cover the costs, so the government needed to know how much to charge the industry. "Can you help?" the official asked (Flyvbjerg and Gardner 2023).

Bent was baffled. At the time, he did not have the data on nuclear decommissioning that we do now. And he did not think he could get any. Very few nuclear power plants had been decommissioned worldwide, and those few had been under exceptional circumstances, such as Chernobyl and Three-Mile Island. Sweden would effectively be the first country in the world to carry out a planned decommissioning of a fleet of reactors. They were truly unique in this sense, it seemed to Bent. "I can't help," he answered. "Sorry."

But the Swedish official saw something Bent did not.

Consultants had given the Swede a report estimating the cost and the "cost risk," meaning the risk of costs higher than expected. He had almost accepted the consultants' estimates, but then he read Flyvbjerg et al. (2003), which documents empirical cost and cost risk for transportation infrastructures like roads, bridges, tunnels, and rail lines. The official noticed that according to this study, the cost risk was higher for that very ordinary sort of transportation infrastructure than for decommissioning his nuclear infrastructure. "That makes no sense," he said. Completing transportation projects takes five to ten years, and people have been building them for centuries. How can it be less risky to decommission nuclear power when it takes much longer, and we have almost no experience doing it?

Bent agreed. That made no sense. The consultants and their report were dropped.

But the official had an idea for a replacement. Why not use Bent's data on cost risk in transportation infrastructure as a "floor" – a minimum – and assume the real cost risk of nuclear decommissioning and storage would be somewhere above that? That would be far from a perfect estimate. But it would be better than the one the consultants had put together. And decommissioning was not starting soon. If the Swedish government put that estimate in place now and got the nuclear industry to start paying into the fund on that basis, the government could adjust the estimate and the payments later as more was learned about decommissioning in Sweden and elsewhere.

Bent was impressed. It was such an elegant and reasonable approach. He made his data available and worked with the official and his team to develop the approach, which became Swedish national policy (Statens Offentlige Utredninger 2004).

The uncomfortable truth is that Bent had fallen prey to uniqueness bias by assuming that a project as unprecedented at the time as nuclear decommissioning had nothing to learn from other projects, including projects he himself had researched, published about, and consulted on. With a little imagination and logic, the Swedish official proved Bent to be wrong. He has never forgotten the lesson. We suggest that neither should you.

Even if many project managers are quick to spontaneously claim their projects as unique, if you give them examples like those above, explain the reasoning regarding local versus global uniqueness, and *then* ask them to come up with examples of truly – i.e., globally – unique projects, they struggle. We have tried this with hundreds of project managers in our leadership training, and so far, no one has been able to give a convincing example of a project that all could agree was truly unique. This conclusion is not just our view but the view of each cohort where we have tried the exercise. The closest we have come was a suggestion at a McKinsey conference for IT leaders where a participant whose company had been involved in the invention and initial rollout of mobile texting suggested the delivery of SMS texting as an example of a truly unique project. The leader explained that it had taken only a few weeks to develop the app for mobile texting, and no one on the project or outside of it had really understood what it was they had invented. Growth in usage was slow at first. The project seemed minor. Nothing to text home about. No one could have predicted the wildfire explosion in usage that would follow, no matter how hard they had tried, and no other project set a precedent for this. So mobile texting was unique in this sense, or so the leader argued, and many of us in the room tended to agree at first.

But then others jumped in and suggested that texting was *not* unprecedented, like NASA did for Apollo above. It is commonplace that demand for a new piece of software cannot be reliably predicted, so there is nothing unique about that, but quite the opposite. A host of communication technologies exist that can be considered precedents to texting, even if we limit ourselves to electronic communications, e.g., the telegraph, the radio, the telephone, the fax machine, and forerunners of today's internet, like ARPANET. A systematic study of the diffusion in time and space of these and other new communication technologies would have given the inventors of texting an idea of the uncertainties and the S-curve growth pattern – with a slow start and acceleration later – that they faced. Had anyone thought about this? No, because they saw texting as both unique *and* unimportant and were, therefore, not motivated to look for similarities. The cohort eventually settled on a consensus that, yes, mobile texting was an unusual project with massive impact, but, no, it was not truly unique. And we concluded, like other cohorts before and after, that we, in fact, could not come up with a single example of a truly unique project. We suggest you try this exercise whenever you or your team begin to think about your projects as unique.

In conclusion, truly unique projects are not just rare. They are exceedingly uncommon, bordering on non-existent. For project management, it does not matter, however, whether truly unique projects exist or not because it is *perceived uniqueness* that counts. Perceived uniqueness not only exists but is pervasive: one in five projects (19.2 percent) in our sample perceived themselves to be highly unique, scoring above seven on our scale. Perceived uniqueness results in uniqueness bias, negatively impacting project performance, as documented above. This is why you should root out perceived uniqueness in your projects if you want them and your business to succeed.

## Direction of Causality

Our findings are based on correlation and regression analyses. As always, such analyses do not prove causality, here from uniqueness bias to project performance. To understand causality, we need to consider reverse causality: decision makers might perceive their project as unique because it underperformed compared to previous experience. Retrospective sensemaking – a form of hindsight bias (Schkade and Kilbourne 1991) – might explain the judgments we captured in our data. Our analysis cannot statistically rule out reverse causality. More research, e.g., a time-lagged panel study, would be needed for this. However, we found statistically highly significant correlations between uniqueness bias and characteristics other than overrun concerning technical aspects of projects (unknowns in the design, complex processing, competing and shifting requirements) as well as aspects of novelty (incremental approach, a novel practice of development), and of political pressure to undertake a project. Such characteristics may be interpreted as antecedents of perceived uniqueness, suggesting that perceived uniqueness is more than retrospective sensemaking of performance.

False perception of uniqueness is a well-established bias in social comparisons where individuals see themselves as more unique than warranted, resulting in biased assessments of risks and opportunities (Weinstein and Klein 1996, Chambers 2008). In management, uniqueness bias can result from managerial tunnel vision, which affects information processing (Posavac et al. 2010). These studies suggest a directional effect from uniqueness bias to project performance, as we suggest for our findings. Similarly, Sanbonmatsu et al. (1991) studied consumer choice between multiple brands and products. Experiments demonstrated that when deciding what to buy, consumers search for unique features instead of common ones among alternatives before making their choice. Bolton (2003) further found that even when decision makers are forced to consider more alternatives, including counterfactuals, decisions are still biased towards the initial scenario, which is typically based on availability bias and the inside view. Again, this supports our interpretation that uniqueness bias is likely the outcome of information processing by decision makers rather than an expression of hindsight bias.

In sum, based on the extant literature and our findings reported above, the evidence strongly suggests that the direction of causality between uniqueness bias and project performance is from the former to the latter.

# How to Curb Uniqueness Bias

People are generally predisposed to uniqueness bias. Project planners and managers perhaps more than most because they have been primed by their professional organizations and the project management literature to think of projects as unique. It is, therefore, no surprise that it takes time and experience to unlearn the predisposition. When we discussed uniqueness bias, a seasoned megaproject manager told us, "The first 20 years as a megaproject manager, I saw uniqueness in each project I did; the next 20 years similarities."

To illustrate what to do in the face of uniqueness bias, consider the chief information officer of a large global logistics company that participated in our research. Project managers from each of the company's flagship projects filled out a questionnaire in which we included questions about uniqueness bias, strategic bias, and the outside view. At the debriefing meeting about the research, the CIO spotted a project scoring themselves as absolutely unique. The CIO wanted to know which project it was. When told it was the installation of a standard software package for supply chain and warehouse automation in the Czech Republic, the CIO said he was surprised because the company had installed and was operating this package for clients in nearly 1,000 other locations. He phoned up the Czech project manager on the spot during the debrief to find out what was going on. The manager explained that the project was unique because it was the first time this product was used in the Czech Republic. He could find no software developers locally who had worked on the software. The CIO responded by saying the project manager should immediately get on a plane to company headquarters to talk to their in-house experts for the software, and he would send him developers from the central IT development team to help with the project. The CIO did right. He saw uniqueness bias for what it is: a fallacy based on localism bias, here in the local Czech context. The project manager's perceived uniqueness did not correspond to actual uniqueness, as usual. Once you understand uniqueness bias, as the CIO did, you know that it undermines project performance and that, therefore, you need to root it out when you encounter it.

Another project leader at a major international bank explained how many of the bank's programs – especially big IT-led change programs – seemed unique at first. But by breaking them down into specific tasks and approaches, it was still possible to bring other teams' experiences from other projects to bear on the seemingly unique projects for those tasks and approaches. For example, the project leader explained that if you are developing a runbook for a go-live migration, you should talk to people who have done migrations before. Or, if you are trying to estimate the lead times in establishing your test environment for a new project, ask other projects and teams for their experience with lead times to get an outside view and use this to challenge the inside view of your team. The project leader had used this approach on projects that were key to the bank's success and described the experience as "invaluable" for avoiding uniqueness bias and underperformance.

Uniqueness bias feeds what Kahneman (2011: 247) calls the inside view. Seeing things from this perspective, planners focus on the specific circumstances of the project they are planning and seek evidence from their own experience. Estimates of budget, schedule, etc., are based on this information, typically built from "the inside and out," or bottom-up, as in conventional cost engineering. We hypothesize that an additional mechanism contributing to underperformance might be that when a project is perceived as unique, the project management is more likely to ignore or circumvent normal processes, e.g., bottom-up estimation, procurement, oversight, and audit, allowing project problems to accumulate undiscovered for longer, leading to cost and time overruns. This is a topic for further research.

The alternative to the inside view is the outside view, which is the cure to uniqueness bias and consists of viewing the project you are planning from the perspective of similar projects that have already been completed, basing your estimates for the planned project on the actual outcomes of these projects. Yet, how decision makers take the outside view is essential in order to avoid reintroducing biases that the approach aims to remove. First, decision makers using personal experience as their source of information tend to underweight rare, catastrophic events (Ludvig et al. 2014, Yechiam et al. 2015). Thus, when taking an outside view, decision makers need to consider the full range of decision outcomes, i.e., the variance in total. Decision makers should not discount catastrophic events, which our analysis shows are statistically significantly associated with decisions that suffer from uniqueness bias. Second, decision makers tend to use anchors, including unsuitable ones (Kahneman and Tversky 1974, Epley and Glovich 2006). The outside view can help correct this by providing a realistic anchor, where realism is measured by agreement with empirical base rates. An anchor, however, has been shown to instill more confidence in decision makers, irrespective of whether anchors are informative or random (Smith and Marshall 2017, Stavrora and Evans 2019). Decision makers should, therefore, use the outside view carefully and critically to correct not only their estimates but also their confidence in those estimates.

If your project were truly unique, then similar projects would not exist, of course, and the outside view would be both impossible and irrelevant. This would leave you with the inside view as the only option for planning your project. If, in contrast, your project is not truly unique, but your team thinks it is, then the outside view will still be left by the wayside, and the inside view will reign supreme. The latter is the typical situation. "In the competition with the inside view, the outside view does not stand a chance," as pithily observed by Kahneman (2011: 249), who further notes that people who have information about an individual case, as you do with the inside view, rarely feel the need to know the statistics of the class to which the case belongs, resulting in base-rate neglect, i.e., playing the casino without knowing the odds. This is obviously not a good idea, and no professional would do it.

Nevertheless, it is common. In Kahneman's example of writing a textbook above, the outside view – represented by the expert – said it would take a minimum of seven years to write the book, with a 40 percent risk of never completing it. In contrast, the inside view told Kahneman and his colleagues that they would write the book with certainty in one and a half to two and a half years. The inside view was hugely wrong. The outside view proved correct. Humans appear to be hardwired for the inside view, which is the perspective people spontaneously adopt when they plan. It is reinforced for project planners and managers by an institutionally amplified uniqueness bias nurtured by their professional organizations and the project management literature. The inside view is, therefore, typical of project planning and management to a degree where it seems appropriate to speak of an "inside-view bias" for this profession, driven by uniqueness bias.

The consequences are dire because only the outside view effectively takes into account all relevant risks, including the so-called "unknown unknowns." These are impossible to predict from the inside because a project can go wrong in too many ways. However, the unknown unknowns are included in the outside view because *anything* that went wrong with the completed projects that constitute the outside view is included in their outcome data (Flyvbjerg 2006). Using these data for planning and managing a new project, therefore, leaves you with a measure of all risks, including unknown unknowns, which is a primary reason why the outside view is more accurate than the inside view. Uniqueness bias makes you blind to unknown unknowns. The outside view is an antidote to such blindness and uniqueness bias.

Fortunately, a range of excellent methods exists for systematically and effectively taking the outside view, eliminating or reducing uniqueness bias. The main methods are reference class forecasting (Flyvbjerg 2006), premortems (Klein 2007), similarity-based forecasting (Lovallo et al. 2012, 2022), and noise audits (Kahneman et al. 2021). Each method achieves the same feat in slightly different ways by introducing the full variance that is relevant to a given decision instead of ignoring variance, especially extreme variance, as is common. Kahneman et al. (2021: 326-28) call this approach "decision hygiene." The antonym to hygiene is "uncleanliness." This is the problem with much decision making: it is unclean – polluted by bias and misinformation – and needs to be cleaned up before informed decisions can be made. The methods for cleaning up are well documented in the literature, to which we refer anyone interested in eliminating uniqueness and other biases from their projects, which is to say, anyone interested in running their projects and organizations successfully. It should be mentioned, however, that even if project management professionals and scholars know about and have these effective and well-documented tools at their disposal, many still, unfortunately, insist on the inside view, as Kahneman predicted they would, and actively reject the tools, often demonstrating a profound lack of understanding of base rates, statistics, and behavioral science (Flyvbjerg et al. 2018, Flyvbjerg 2018).

## Conclusions

In this study, we explored experts' real-world decisions to invest time, money, and effort in projects. Such decisions are based on forecasts of budgets, timelines, and other resource requirements. We explored uniqueness bias – the tendency of planners and managers to see their projects as singular – as a possible explanation of why, with alarming frequency, forecasts turn out to be highly inaccurate and decisions – in hindsight – not what they should have been. We found a statistically highly significant association between projects' perceived uniqueness and underperformance. Projects considered truly unique have a cost overrun that is 45 percentage points higher in real terms than projects considered not unique, on average. Furthermore, truly unique projects have a 37 percent risk of extreme cost overrun, compared with 1 percent for not-unique-at-all projects.

The existence of uniqueness bias in project planning and management is unsurprising. The very definitions of projects and project management endorsed by standard-setting bodies like the PMI and much of the project management literature emphasize uniqueness and thereby encourage uniqueness bias, undermining project performance. Uniqueness bias has been described before in individuals (Suls and Wan 1987, Suls et al. 1988, Goethals et al. 1991, Rose et al. 2017) and linked to comparative optimism (Rose et al. 2017). The bias has also been suspected to occur in professional, expert decision making, and in organizations (Flyvbjerg 2014, 2021). For the first time, the present study documents that (a) uniqueness bias is indeed found with experts and organizations and (b) here the bias is statistically significantly associated with error in forecasting and underperformance of investment projects.

To curb uniqueness bias, decision makers need to re-frame their decisions. First, we suggest the term "unique" be simply dropped from the vocabulary of decision making and from definitions of projects and project management. Second, we propose consistent and widespread use of what Kahneman et al. (2021) felicitously call "decision hygiene," for instance, through reference class forecasting, premortems, and noise audits.

It is easy to understand why people would think their decisions and projects are unique. It is "fast" thinking, according to Kahneman's (2011) famous term, which is the default mode of how humans think. As such, it also compromises thinking, as documented by Kahneman, Tversky, and other behavioral scientists. It saves project planners and managers the considerable effort of figuring out to which class of project a planned project belongs, what the base rates and tails are for that class, how those rates and tails translate into risk, and, finally, how the risk may be mitigated if needed. If your project is unique, it self-evidently does not belong to any class of project, so you can skip the slow and cumbersome work of answering all these questions.

But "fast" thinking is sloppy thinking in the context of project investment decision making. The problem is that most, if not all, things are "one of something," i.e., they are not unique. As explained in Flyvbjerg and Gardner (2023), you need precisely Kahneman's slow thinking to figure out what that "something" is and what it means to your project and organization in terms of risk. The first step to delivering successful projects is, therefore, to unlearn the conventional definitions of projects and project management and realize that your project is *not* unique. It is "one of those."

# Methodological Appendix

Our analysis included a list of 36 explanatory variables of extreme cost overrun, captured in interviews with project leaders and their teams and shown in Table 2. We used the list to build a parsimonious model by excluding all variables that had no statistically significant association with either (a) extreme cost overrun (coded as 0 or 1 and measured by using Tukey's definition of outliers, see main text) or (b) perceived uniqueness, arriving at a short list of variables making up the model shown in Figure 4. The model proved to have an adequate overall fit, with a comparative fit index (CFI) of 0.921 (near the conventional cut-off at 0.95) and a Tucker-Lewis Index (TLI) of 0.861. The root mean square error of approximation (RMSEA) is 0.076 (near the conventional cut-off at 0.05). The standardized root mean square residual (SRMR) is 0.054 (below the conventional cut-off at 0.08).

In the model, perceived uniqueness serves as a mediator of the mix of internal and external project resources, with a direct and an indirect effect. Projects perceived as more unique have a less sustainable mix of internal and external resources, meaning they rely more heavily on external input. The total indirect effect via the moderator is negative, which means that having an unsustainable mix of internal and external resources increases the likelihood of extreme cost overrun, i.e., a cost overrun outlier. However, while the pathways in the model individually are statistically significant, the total indirect effect is not (p = 0.052), as shown in Table 3 below.

*Table 2: List of 36 explanatory variables of extreme cost overrun, used as the basis for the parsimonious model.*

| Success Factor Domain | Variable Number | Full Statement | Shortened Label |
|---|---|---|---|
| Project complexity | 1 | The goals, business case, and success criteria were clear. | Clear goals and business case |
| | 2 | The project had competing and shifting requirements. | Stable requirements |
| | 3 | The project could influence its environment in order to achieve success. | Environmental influence |
| | 4 | The project involved innovative technology. | Innovation |
| | 5 | There were a large number of unknowns in the solution design. | Few unknowns in the design |
| | 6 | The project followed an incremental approach. | Incremental approach |
| | 7 | The deliverables of the project were highly inter-dependent. | Decoupled deliverables |
| | 8 | The scope was minimal and stable. | Minimal and stable scope |
| | 9 | The project faced resistance in the organization. | No resistance |
| | 10 | The project had an effective governance structure. | Effective governance |
| | 11 | Senior management supported the project. | Sr. mgmt. support |
| Project management | 12 | Effectively monitored and controlled, e.g., by a PMO? | Monitoring and control |
| | 13 | Employing robust supplier contracts with clear responsibilities? | Robust contracting |
| | 14 | Using a sustainable mix of internal and external resources? | Sustainable int-ext mix |
| | 15 | Managing change effectively? | Effective change mgmt |
| | 16 | Addressing and managing risks effectively? | Effective risk mgmt |
| | 17 | Involving users? | User involvement |
| | 18 | Using an effective management methodology? | Effective PM method |
| | 19 | Having an effective project sponsor? | Effective sponsor |
| | 20 | Able to align the major stakeholders? | Aligned stakeholders |
| | 21 | Transparently communicating the project status? | Transparent status communication |
| | 22 | Lead by competent project manager(s)? | Competent project managers |
| | 23 | Carried out by a qualified and motivated team? | Qualified and motivated team |
| | 24 | Planned in detail and the plan updated regularly? | Detailed, up-to-date plan |
| | 25 | Following a realistic and reliable schedule? | Realistic, reliable schedule |
| | 26 | Lacking suitable staff? | Suitably staffed |
| | 27 | Good in communication among team members? | Good team communication |
| | 28 | Supported by well-performing suppliers? | Well performing supplier |
| Planning | 29 | This project is unique and therefore it is difficult to compare with other projects. | Uniqueness |
| | 30 | The initial plan and business case showed political intent instead of the most likely outcome | No political intent |
| | 31 | If we had known the real costs of the project, it would not have been built | No regrets |
| | 32 | The organizational unit has a historical record of past projects that this business case was compared to | Outside view |
| Technical complexity | 33 | The solution requires complex internal processing in the system | Low processing requirements |
| | 34 | The solution needs to be reusable | Low reusability requirements |
| | 35 | The solution needs to fulfil high-performance requirements | Low performance requirements |
| | 36 | Security aspects are very important for the solution | Low security requirements |

*Figure 4: Parsimonious model of project characteristics that have a statistically significant association with perceived uniqueness and with the project being a Tukey outlier in terms of cost overrun.*

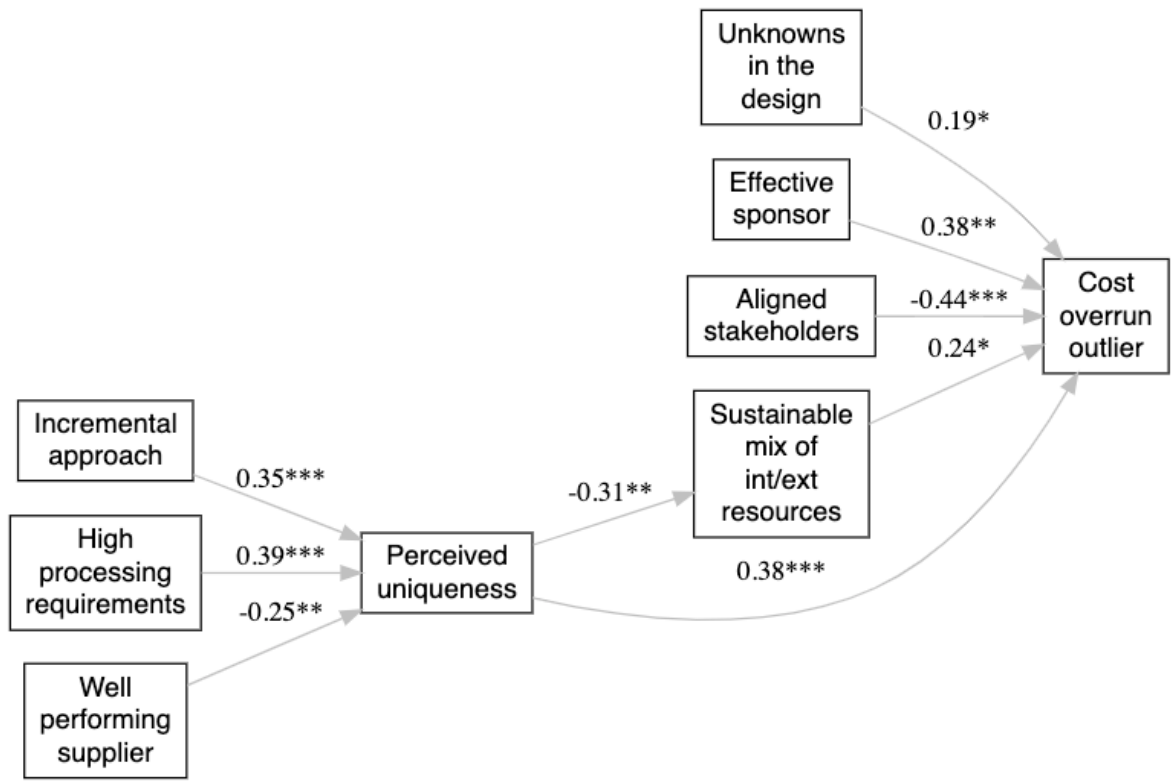

The model shows that (a) the number of unknowns in a project's solution design and (b) misaligned stakeholders increase the likelihood of a project having extreme cost overrun. Interestingly, the model further shows that having an effective project sponsor also increases the likelihood of extreme overrun. This might seem counterintuitive at first, but it aligns with previous research, which indicates that effective sponsors are less likely to abandon their projects, even when they should have. In the present model, this factor suggests that the effective project sponsor likely would have avoided the termination of their projects.

Finally, perceived uniqueness has three antecedents in the model: an incremental approach, high processing requirements, and a well-performing supplier. Taking a more incremental approach, which here means an agile approach, correlates with the increased uniqueness of the project. So do high technical processing requirements, e.g., real-time processing and low latency. A well-performing supplier is an antidote to perceived uniqueness, providing evidence that engaging with this type of supplier may well be worth your while if you are a project leader or manager.

*Table 3: Structural equation model estimates for the parsimonious model of project characteristics associated with perceived uniqueness and cost overrun outliers (standardized estimates based on variances of observed and latent variables).*

| Regressions | Estimate | Standard error | z-value | p-value | Standardized estimate |
|---|---|---|---|---|---|
| *Cost overrun outlier* | | | | | |
| Perceived uniqueness | 0.05 | 0.01 | 3.94 | 0.0000 | 0.38 |
| Sustainable mix of int/ext resources | 0.04 | 0.01 | 2.56 | 0.0100 | 0.24 |
| Unknowns in the design | 0.03 | 0.01 | 2.06 | 0.0390 | 0.19 |
| Effective sponsor | 0.07 | 0.02 | 3.02 | 0.0030 | 0.38 |
| Aligned stakeholder | -0.08 | 0.02 | -3.49 | 0.0000 | -0.44 |
| *Perceived uniqueness* | | | | | |
| Incremental approach | 0.32 | 0.08 | 3.95 | 0.0000 | 0.35 |
| High processing requirements | 0.40 | 0.09 | 4.46 | 0.0000 | 0.39 |
| Well performing suppliers | -0.29 | 0.10 | -2.86 | 0.0040 | -0.25 |
| *Sustainable mix of int/ext resources* | | | | | |
| Perceived uniqueness | -0.26 | 0.09 | -2.99 | 0.0030 | -0.31 |
| *Moderation of perceived uniqueness* | | | | | |
| Indirect effect | -0.01 | 0.01 | -1.95 | 0.0520 | -0.07 |
| Total effect | 0.04 | 0.01 | 3.21 | 0.0010 | 0.30 |
| Variances | | | | | |
| Cost overrun outliers | 0.08 | 0.01 | 6.44 | 0.0000 | 0.65 |
| Sustainable mix of int/ext resources | 4.68 | 0.73 | 6.44 | 0.0000 | 0.90 |
| Perceived uniqueness | 4.87 | 0.76 | 6.44 | 0.0000 | 0.64 |
| R-Square | | | | | |
| Cost overrun outliers | 0.35 | | | | |
| Sustainable mix of int/ext resources | 0.10 | | | | |
| Perceived uniqueness | 0.36 | | | | |

The model can be further simplified and improved by removing the mediator relationship because (a) the indirect effect was not statistically significant (p = 0.052, as discussed above) and (b) the unknowns in the design were not highly statistically significant (p = 0.039), see Figure 5 and model estimates in Table 4.

After this adaption, the model fit is fully acceptable: The comparative fit index (CFI) is 0.978, which is above the conventional cut-off at 0.95; the Tucker-Lewis Index (TLI) is 0.951. The root mean square error of approximation (RMSEA) is 0.043, which is below the conventional cut-off at 0.05. Finally, the standardized root mean square residual (SRMR) is 0.034, which is below the conventional cut-off of 0.08.

The model now shows that the likelihood of a cost outlier is associated with three project characteristics: perceived uniqueness, aligned stakeholders (negative association), and effective sponsor. As discussed above, the counterintuitive relationship with the effectiveness of a sponsor might indicate that a project was not terminated or cancelled as – one could argue – it should have been. Perceived uniqueness, in turn, is associated with three project characteristics: an incremental approach to the project, high processing requirements, and a well-performing supplier (negative association).

*Figure 5: Further simplified and improved parsimonious model.*

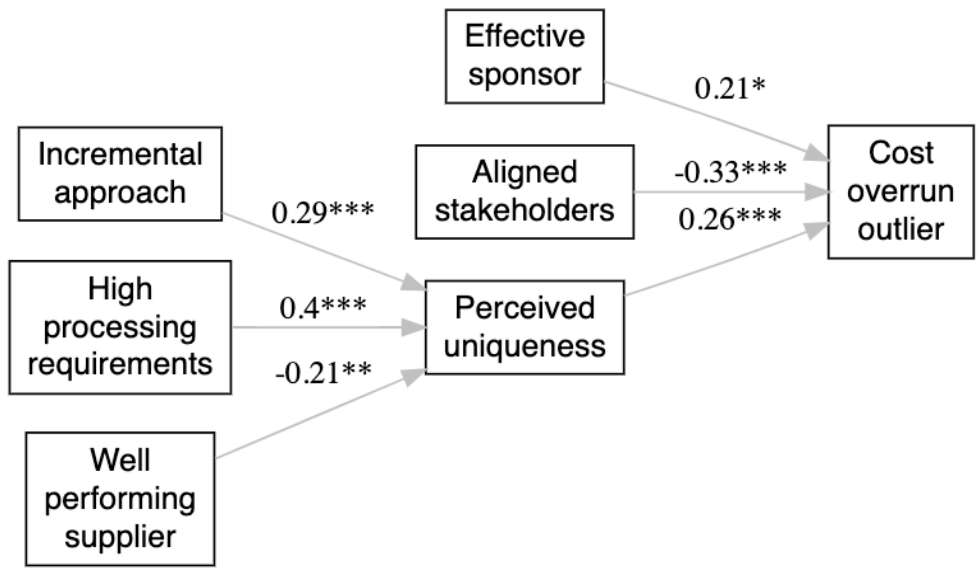

*Table 4: Structural equation model estimates for the further simplified and improved parsimonious model (standardized estimates based on observed and latent variables).*

| Regressions | Estimate | Standard error | z-value | p-value | Standardized estimate |
|---|---|---|---|---|---|
| *Cost overrun outlier* | | | | | |
| Perceived uniqueness | 0.03 | 0.01 | 3.51 | 0.0000 | 0.26 |
| Effective sponsor | 0.03 | 0.02 | 2.22 | 0.0270 | 0.21 |
| Aligned stakeholder | -0.05 | 0.01 | -3.53 | 0.0000 | -0.33 |
| *Perceived uniqueness* | | | | | |
| Incremental approach | 0.23 | 0.05 | 4.20 | 0.0000 | 0.29 |
| High processing requirements | 0.40 | 0.07 | 5.80 | 0.0000 | 0.40 |
| Well performing suppliers | -0.27 | 0.09 | -3.13 | 0.0020 | -0.21 |
| Variances | | | | | |
| Cost overrun outliers | 0.06 | 0.01 | 8.97 | 0.0000 | 0.86 |
| Perceived uniqueness | 4.71 | 0.53 | 8.97 | 0.0000 | 0.75 |
| R-Square | | | | | |
| Cost overrun outliers | 0.14 | | | | |
| Perceived uniqueness | 0.25 | | | | |

# Division of Work

Bent Flyvbjerg identified uniqueness bias as relevant not only to individuals, as in conventional behavioral science, but also to projects and organizations; he got the idea for the paper, developed its main concepts, did the research, and wrote the initial drafts. Alexander Budzier contributed to the initial drafts and co-authored subsequent revisions; he also provided help with the statistical study reported in the section "Does Uniqueness Bias Impact Project Performance?" Specifically, he designed the survey, collected the data, conducted the first statistical analyses, created graphics, and helped interpret results. M. D. Christodoulou and M. Zottoli did additional statistical analyses and, as the professional statisticians on the team, did quality control of all statistical analyses in the paper.

# References

APM, Association of Project Management (2012). *APM Body of Knowledge*, 6th Edition (High Wycombe: Association for Project Management).

APM, Association of Project Management (2019). *APM Body of Knowledge*, 7th Edition (High Wycombe: Association for Project Management).

Baccarini, D. (2004). Modelling Complex Projects, *International Journal of Project Management*, *22*(6), 519–520.

Bazerman, M. H. (2001). The study of "real" decision making. *Journal of Behavioral Decision Making*, *14*(5), 353.

Blanchard, M. D., Jackson, S. A., & Kleitman, S. (2020). Collective decision making reduces metacognitive control and increases error rates, particularly for overconfident individuals. *Journal of Behavioral Decision Making*, *33*(3), 348-375.

Bolton, L. E. (2003). Stickier priors: The effects of nonanalytic versus analytic thinking in new product forecasting. *Journal of Marketing Research*, *40*(1), 65-79.

British Olympic Association, 2000, *London Olympic Bid: Confidential Draft Report to Government*, 15 December (London: British Olympic Association Library).

Budzier, A. and Bent Flyvbjerg, 2024, "The Oxford Olympics Study 2024: Are Cost and Cost Overrun at the Games Coming Down?" *Saïd Business School Working Paper*, University of Oxford, 27 pp.

Buehler, R., Messervey, D., & Griffin, D. (2005). Collaborative planning and prediction: Does group discussion affect optimistic biases in time estimation?. *Organizational Behavior and Human Decision Processes*, *97*(1), 47-63.

Chambers, J. R. (2008). Explaining false uniqueness: Why we are both better and worse than others. *Social and Personality Psychology Compass*, *2*(2), 878-894.

Chapman, R. J. (1998). The role of system dynamics in understanding the impact of changes to key project personnel on design production within construction projects. *International Journal of Project Management*, *16*(4), 235-247.

Conroy, G., & Soltan, H. (1998). ConSERV, a project specific risk management concept. *International Journal of Project Management*, *16*(6), 353-366.

Czuchry, A. J., & Yasin, M. M. (2003). Managing the project management process. *Industrial Management & Data Systems*, *103*(1), 39-46.

Dannals, J. E., & Oppenheimer, D. M. How people deal with…............................. outliers. *Journal of Behavioral Decision Making*, e2303.

Epley, N., & Gilovich, T. (2006). The anchoring-and-adjustment heuristic: Why the adjustments are insufficient. *Psychological science*, *17*(4), 311-318.

Feduzi, A., & Runde, J. (2014). Uncovering unknown unknowns: Towards a Baconian approach to management decision-making. *Organizational Behavior and Human Decision Processes*, *124*(2), 268-283.

Fiedler, K., Hütter, M., Schott, M., & Kutzner, F. (2019). Metacognitive myopia and the overutilization of misleading advice. *Journal of Behavioral Decision Making*, *32*(3), 317-333.

Flyvbjerg, B. (2006). From Nobel Prize to Project Management: Getting Risks Right, *Project Management Journal*, *37*(3), 5-15.

Flyvbjerg, B. (2014). What You Should Know about Megaprojects and Why: An Overview, *Project Management Journal*, *45*(2), 6-19.

Flyvbjerg, B. (2016). The Fallacy of Beneficial Ignorance: A Test of Hirschman's Hiding Hand, *World Development*, *84*(May), 176–189.

Flyvbjerg, B. (2018). Planning Fallacy or Hiding Hand: Which Is the Better Explanation? *World Development*, *103*, 383-386.

Flyvbjerg, B. (2021). Top Ten Behavioral Biases in Project Management: An Overview, *Project Management Journal*, *52*(6), 531–546.

Flyvbjerg, B., Ansar, A., Budzier, A., Buhl, S., Cantarelli, C., Garbuio, M., Glenting, C., Skamris Holm, M., Lovallo, D., Lunn, D., Molin, E., Rønnest, A., Stewart, A., & van Wee, B. (2018). Five Things You Should Know about Cost Overrun, *Transportation Research Part A: Policy and Practice*, *118*(December), 174-190,

Flyvbjerg, B., Bruzelius, N., & Rothengatter, W. (2003). *Megaprojects and Risk: An Anatomy of Ambition* (Cambridge University Press).

Flyvbjerg, B., & Budzier, A. (2011). Why Your IT Project May Be Riskier than You Think, *Harvard Business Review*, *89*(9), 23-25.

Flyvbjerg, B., Budzier, A., Lee, J. S., Keil, M., Lunn, D., & Bester, D. W. (2022). The Empirical Reality of IT Project Cost Overruns: Discovering A Power-Law Distribution. *Journal of Management Information Systems*, *39*(3), 607-639.

Flyvbjerg, B., Budzier, A., & Lunn, D. (2021). Regression to the tail: Why the Olympics blow up. *Environment and Planning A: Economy and Space*, *53*(2), 233-260.

Flyvbjerg, B., & Gardner, D. (2023). *How Big Things Get Done: The Surprising Factors that Determine the Fate of Any Project from Home Renovations to Space Exploration, and Everything in Between* (New York: Penguin Random House).

Fox, J. R., & Miller, D. B. (2006). *Challenges in Managing Large Projects* (Fort Belvoir, VA: Defense Acquisition University Press).

Gaddis, P. O. (1959). The Project Manager, *Harvard Business Review*, *37*(3), 89–97.

Gareis, R. (1991). Management by projects: the management strategy of the "new" project-oriented company. *International Journal of Project Management*, *9*(2), 71-76.

Gigerenzer, G. (2018). The Bias Bias in Behavioral Economics, *Review of Behavioral Economics*, *5*, 303-336.

Goethals, G. R., Messick, D. M., & Allison, S. (1991) The Uniqueness Bias: Studies in Constructive Social Comparison, in Suls, J. and Wills, T. A. (Eds.). *Social Comparison: Contemporary Theory and Research* (Hillsdale, NJ: Erlbaum), 149-176.

Grün, O. (2004). *Taming Giant Projects: Management of Multi-Organization Enterprises* (Berlin and New York: Springer).

Hirschman, A. O. (2014). *Development Projects Observed*, third edition published as a Brookings Classic with a new foreword by Cass R. Sunstein and a new afterword by Michele Alacevich; first published 1967.

Jaafari, A. (2007). Project and program diagnostics: A systemic approach. *International journal of project management*, *25*(8), 781-790.

Janis, I. L. (1983). *Groupthink* (Boston: Houghton Mifflin).

Jennings, W. (2012). *Olympic Risks* (New York, London: Palgrave Macmillian).

Kahneman, D. (2011). *Thinking, Fast and Slow* (New York: Farrar, Straus and Giroux).

Kahneman, D., & Klein, G. (2009). Conditions for Intuitive Expertise: a Failure to Disagree, *American Psychologist*, *64*(6), 515-526.

Kahneman, D., Sibony, O., & Sunstein, C. R. (2021). *Noise: A Flaw in Human Judgment* (London: William Collins).

Kahneman, D., & Tversky, A. (1972). Subjective probability: A judgment of representativeness. *Cognitive psychology*, *3*(3), 430-454.

Kahneman, D., & Tversky, A. (1977). *Intuitive prediction: Biases and corrective procedures*. Decisions and Designs Inc Mclean Va.

Kirshner, S. N. (2021). Construal level theory and risky decision making following near-miss events. *Journal of Behavioral Decision Making*, *34*(3), 379-392.

Klein, G. (2007). Performing a Project Premortem, *Harvard Business Review*, September, 1-2.

Leisti, T., & Häkkinen, J. (2018). Learning to decide with and without reasoning: How task experience affects attribute weighting and preference stability. *Journal of Behavioral Decision Making*, *31*(3), 367-379.

Lovallo, D., Clarke, C., & Camerer, C. (2012). Robust Analogizing and the Outside View: Two Empirical Tests of Case-Based Decision Making, *Strategic Management Journal*, *33*, 496–512.

Lovallo, D., Cristofaro, M., & Flyvbjerg, B. (2022). Addressing Governance Errors and Lies in Project Forecasting, forthcoming in *Academy of Management Perspectives*.

Ludvig, E. A., Madan, C. R., & Spetch, M. L. (2014). Extreme outcomes sway risky decisions from experience. *Journal of Behavioral Decision Making*, *27*(2), 146-156.

Magee, J. F. (1982). Management Challenges in the 1980s: What Is Ahead in Project Management, in Kelley, A. J. (Ed.) *New Dimensions of Project Management* (Lexington, MA: Lexington Books).

Mandelbrot, B. B. (1997). *Fractals and Scaling in Finance*, New York: Springer.

Merrow, E. W. (2011). *Industrial Megaprojects: Concepts, Strategies, and Practices for Success* (Hoboken, NJ: Wiley).

Metcalfe, J. (1998). Cognitive optimism: Self-deception or memory-based processing heuristics?. *Personality and Social Psychology Review*, *2*(2), 100-110.

Moore, D. A., & Healy, P. J. (2008). The trouble with overconfidence. *Psychological review*, *115*(2), 502.

Paine, J. W., Federspiel, F., Seifert, M., & Zou, X. (2020). Take a risk or proceed with caution: Prevention motivation moderates responses to near-loss events. *Journal of Behavioral Decision Making*, *33*(4), 505-522.

Perminova, O., Gustafsson, M., & Wikström, K., 2008. "Defining uncertainty in projects–a new perspective." *International Journal of Project Management*, *26*(1), 73-79.

Plonsky, O., & Teodorescu, K. (2020). The influence of biased exposure to forgone outcomes. *Journal of Behavioral Decision Making*, *33*(3), 393-407.

PMI, Project Management Institute (2021). *A Guide to the Project Management Body of Knowledge (PMBOK® Guide)*, 7th Edition (Newtown Square, PA: Project Management Institute, Inc.).

Posavac, S. S., Kardes, F. R., & Brakus, J. J. (2010). Focus induced tunnel vision in managerial judgment and decision making: The peril and the antidote. *Organizational Behavior and Human Decision Processes*, *113*(2), 102-111.

Radzevick, J. R., & Moore, D. A. (2008). Myopic biases in competitions. *Organizational Behavior and Human Decision Processes*, *107*(2), 206-218.

Ramirez, J. E. (2021). *Toward a Theory of Behavioral Project Management*, doctoral dissertation (Chicago: The Chicago School of Professional Psychology).

Read, D., & Grushka-Cockayne, Y. (2011). The similarity heuristic. *Journal of Behavioral Decision Making*, *24*(1), 23-46.

Rose, J. P., Aspiras, O., Vogel, E., Haught, H., & Roberts, L. (2017). Comparative optimism and event skewness. *Journal of Behavioral Decision Making*, *30*(2), 236-255.

Schkade, D. A., & Kilbourne, L. M. (1991). Expectation-outcome consistency and hindsight bias. *Organizational Behavior and Human Decision Processes*, *49*(1), 105-123.

Segelod, E. (2018). *Project Cost Overrun: Causes, Consequences, and Investment Decisions*, Cambridge: Cambridge University Press.

Smith, A. R., & Marshall, L. D. (2017). Confidently biased: comparisons with anchors bias estimates and increase confidence. *Journal of Behavioral Decision Making*, *30*(3), 731-743.

Statens Offentlige Utredninger, SOU (2004). *Betalningsansvaret för kärnavfallet*, SOU 2004:125; Stockholm: Statens Offentlige Utredninger

Stavrova, O., & Evans, A. M. (2019). Examining the trade-off between confidence and optimism in future forecasts. *Journal of Behavioral Decision Making*, *32*(1), 3-14.

Suls, J., & Wan, C. K. (1987). In Search of the False Uniqueness Phenomenon: Fear and Estimates of Social Consensus, Journal of Personality and Social Psychology, 52, 211-217.

Suls, J., Wan, C. K., & Sanders, G. S. (1988). False Consensus and False Uniqueness in Estimating the Prevalence of Health-Protective Behaviors, Journal of Applied Social Psychology, 18, 66-79.

Taleb, N. N. (2007). The Black Swan: The Impact of the Highly Improbable (London and New York: Penguin).

Taleb, N. N. (2020). Statistical Consequences of Fat Tails: Real World Preasymptotics, Epistemology, and Applications, Technical Incerto Collection (New York: STEM Academic Press).

Toplak, M. E., West, R. F., & Stanovich, K. E. (2017). Real-world correlates of performance on heuristics and biases tasks in a community sample. Journal of Behavioral Decision Making, 30(2), 541-554.

Turner, J. R., & Müller, R. (2003). On the Nature of the Project as a Temporary Organization, International Journal of Project Management, 21, 1–8.

Webb, J. (1969). Space-Age Management: The Large-Scale Approach (New York: McGraw-Hill).

Weinstein, N. D., & Klein, W. M. (1996). Unrealistic optimism: Present and future. Journal of Social and Clinical Psychology, 15(1), 1-8.

Yechiam, E., Rakow, T., & Newell, B. R. (2015). Super-underweighting of rare events with repeated descriptive summaries. *Journal of Behavioral Decision Making*, *28*(1), 67-75.